# Predictability of IL-28B-polymorphism on protease-inhibitor-based triple-therapy in chronic HCV-genotype-1 patients: A meta-analysis

Nicolae-Catalin Mechie, Christian Röver, Silke Cameron, Ahmad Amanzada

**Nicolae-Catalin Mechie, Silke Cameron, Ahmad Amanzada,** Department of Gastroenterology and Endocrinology, University Medical Center Goettingen, Georg-August-University, 37075 Goettingen, Germany

**Christian Röver,** Department of Medical Statistics, University Medical Center Goettingen, Georg-August-University, 37073 Goettingen, Germany

**Abstract**

**AIM:** To investigate the predictability of interleukin-28B single nucleotide polymorphism rs12979860 with respect to sustained virological response (SVR) in chronically hepatitis C virus (HCV) genotype-1 patients treated with a protease-inhibitor and pegylated interferon-$\alpha$ (Peg-INF-$\alpha$) based triple-therapy.

**METHODS**: We searched PubMed, the Cochrane Library and Web of Knowledge for studies regarding the interleukin 28B (IL-28B)-genotype and protease-inhibitor based triple-therapy. Ten studies with 2707 patients were included into this meta-analysis. We used regression methods in order to investigate determinants of SVR.

**RESULTS**: IL-28B-CC-genotype patients achieved higher SVR rates (odds 5.34, CI: 3.81-7.49) than IL-28B-non-CC-genotype patients (1.88, CI: 1.43-2.48) receiving triple-therapy. The line of therapy (treatment-naïve or -experienced for Peg-INF-$\alpha$) did not affect the predictive value of IL-28B ($P$ = 0.1). IL-28B-CC-genotype patients treated with protease inhibitor-based triple-therapy consisting of Boceprevir, Simeprevir, Telaprevir or Vaniprevir showed odds of 1.86, 9.77, 4.51 and 0.89, respectively. The *odds* for CC genotype patients treated with Faldaprevir cannot be quantified, as only a single study with a 100% SVR rate was available.

**CONCLUSION**: IL-28B-SNP predicts the outcome for chronic HCV genotype-1 patients receiving protease inhibitor-based triple-therapy. The predictive value varies between the different protease inhibitors.

**Core tip:** Hepatitis C is a world health problem and represents a dynamic field of research for new therapeutic options. Recently direct antiviral agents such as protease inhibitors have been developed which, in addition to pegylated interferon-$\alpha$ and Ribavirin, obtain higher sustained virological response (SVR) rates. Of note, costs are higher and side effects are more common. The data regarding the predictive value of Interleukin 28B (IL-28B) are controversial. This meta-analysis was conducted on 2707 patients treated with different protease inhibitors. Its aim was to clarify the predictive value of IL-28B on SVR in protease inhibitor-based triple-therapy, allowing the possibility of personalized treatment.

# INTRODUCTION

Hepatitis C virus (HCV) is a global health Problem. According to the World Health Organization, approximately 150 million people are chronically infected with HCV, and it is estimated that more than 350 thousand are dying each year[1]. HCV is responsible in Europe and North America for 50% of liver cirrhosis and 25% of hepatocellular carcinoma[2-4].

HCV has 7 genotypes (1 to 7) and approximately 100 subtypes[5]. Genotype 1, which is the most common HCV genotype in Western countries, has the worst prognosis and response to antiviral treatment in comparison to other genotypes[6-8].

In the last years the standard therapy (Standard of care, SOC) for HCV consisted of pegylated Interferon-$\alpha$ (Peg-IFN-$\alpha$) and Ribavirin (RBV)[9]. Recently, several direct antiviral agents (DAA) were developed, such as the protease inhibitors (PI) Boceprevir (BOC), Telaprevir (TVR), Vaniprevir (VNP), Faldaprevir (FLP) and Simeprevir (SMP)[10-13]. In pivotal studies, patients treated with BOC or TVR and Peg-IFN-$\alpha$/RBV achieved significantly higher sustained virological response (SVR) rates compared to standard therapy[11-14]. These new treatment options bring new hopes for chronically HCV infected patients but they have more side effects and higher costs[10].

Treatment predictors are important tools for the management of therapy in patients with chronic HCV infection. For the current standard treatment with Peg-IFN-$\alpha$/RBV in patients with chronic HCV infection, HCV genotypes 2 and 3, low baseline viral load, ethnicity, younger age, low $\gamma$-GT levels, low $\gamma$-GT/ALT level, absence of advanced fibrosis/cirrhosis, and absence of steatosis in the liver have been identified as independent pretreatment predictors of a SVR[15,16].

After initiation of treatment, rapid virological response (RVR, undetectable HCV-RNA at week 4 of therapy) is the best predictor of SVR independent of HCV genotype[16]. Recently, several genome-wide association studies showed that a single nucleotide polymorphism (SNP) within the interleukin 28B (IL-28B) gene is significantly associated with treatment outcome under standard treatment in chronically HCV genotype-1 infected patients[17-19]. IL-28B rs12979860 is the most investigated allele of IL-28B in Europe and North America. The data about the predictive value of IL-28B-genotype in HCV genotype-1 and triple-therapy are inconsistent. In the studies with VNP, IL-28B-genotype had no predictive value for the treatment[13,20]. In the studies by Poordad et al[21], Fried et al[22], Bronowicki et al[23], Sulkowski et al[24] and Akuta et al[25], IL-28B-CC-genotype had a favorable prognosis. In the study by Flamm et al[26] for Boceprevir, genotype IL-28B-TT had a favorable prognosis and by Jacobson et al[27] and Pol et al[28], IL-28B-genotype had a limited influence on SVR. However, more information about the predictability of IL-28B-genotype would allow physicians to individualize antiviral HCV therapy.

Therefore, we conducted this meta-analysis to investigate the predictive value of IL-28B rs12979860 (CC vs CT + TT) allele for SVR in chronically HCV genotype-1 infected patients treated with a triple-therapy regimen consisting of a DAA (BOC, TVR VNP, FLP or SMP) and Peg-IFN-α/RBV.

**MATERIALS AND METHODS**
We searched in PubMed, Web of Knowledge and the Cochrane Library databases, for relevant articles (full text and meeting abstracts) up to January 2014 regarding the following the next key words: "Boceprevir" or/and "SCH503034", "Telaprevir" or/and "VX-950", "Ciluprevir" or/and "BILN 2061", "Simeprevir" or/and "TMC435", "Danoprevir" or/and "R7227", "Vaniprevir" ("MK-7009"), "MK-5172", "Faldaprevir" ("BI201335"), "Narlaprevir" ("SCH900518"), "Asunaprevir" ("BMS-650032"), "PHX1766", "GS-9256", "GS-9451", "ABT450", "IDX320", "ACH-1625". All these DAAs were used as search words in order to avoid missing studies which have determined IL-28B polymorphism for a triple therapy. Because a large number of patient samples were retrospectively tested for IL-28B genotype and some of these results were only presented in meetings, we have decided to include also the meeting abstracts in our meta-analysis. In order to identify relevant studies, the references of the articles included were manually searched. We did not find any other articles that corresponded to our inclusion criteria. The studies search was performed using manual search for Cochrane Library and EndNote X7 for PubMed and Web of Knowledge databases.

The inclusion criteria were: studies with human subjects, more than 18 years of age, HCV genotype-1 patients, treatment with triple-therapy (IFN therapy-naïve and –experienced) with determined IL-28B genetic polymorphism for rs12979860 allele. Only articles in English were included. The exclusion criteria were: HCV/HIV or HCV/HBV co-infection, liver transplantation recipients, pediatric studies and IL-28B genetic polymorphism other than rs12979860. SVR was defined as undetectable HCV-RNA 24 wk after end of treatment.

The studies were reviewed independently by two authors (NCM and AA). All differences were resolved by consensus among these two authors. Our analysis was based on the original published data. For consistency we refrained from contacting the authors of the individual studies. From the studies, the following data were extracted: First author, year of publication, type of patients (IFN therapy-naïve or -experienced), total number of patients, the number of patients with determined IL-28B-genotype, type of DAA, IL-28B genetic polymorphism.

The statistical analysis was performed by CR. We used logistic regression to model the chance of a SVR and investigate potential influential factors. In a logistic regression, binary outcome data are modeled based on the *odds* of events (here: SVR). As is usual regression, the *odds* are then formulated as a function of (potential) explanatory variables. Random effects were included in order to accommodate heterogeneity between studies[29]. As the available data allow to fit a multitude of plausible variations of regression models to the data, we approached the *model selection* problem *via* Bayesian Information Criterion (BIC)[30], which allows to compare and select models based on a single adequacy measure. All analyses were performed using the *R* software (www.r-project.org) and the *lme4* package.

**RESULTS**

*Literature search*

Four thousand three hundred and thirty-seven studies were initially identified on the bases of DAAs. After removing duplicate citations, the remaining 1522 studies were searched for data regarding IL-28B polymorphism and qualified for abstract review. Among the remaining studies, 1454 studies had no data regarding IL-28B and were excluded. The rest 68 studies were selected for a "full paper review". Among these remaining 68 studies, five of them were reviews. Four of them included only interferon-free therapy. There were three meta-analyses which were excluded. Five studies described only SOC therapy. Another 41 studies and meeting abstracts, including preliminary and subgroup analysis from large trials data, rs8099917 IL-28B allele and non-human studies, had to be excluded (Figure 1).

This meta-analysis is based on the following 10 studies: 7 full text studies and 3 meeting abstracts with a total of 2707 IL-28B patients. The studies of Akuta *et al*[25], Bronowicki *et al*[23], Jacobson *et al*[27] and Pol *et al*[28] investigated the interaction between IL-28B genotype and SVR in patients receiving TVR based triple-therapy. The study from Akuta *et al*[25] had no patients with IL-28B genotype receiving SOC. The studies of Flamm *et al*[26] and Poordad *et al*[21] analyzed the BOC based triple-therapy. VNP was used as DAA in the studies of Lawitz *et al*[20] and Manns *et al*[13]. For SMP and FLP only one study could be included for each of them (Fried *et al*[22] and Sulkowski *et al*[24]; Table 1).

***Comparison of dual and triple therapy***

Figure 2 illustrates the estimated *odds* and associated *confidence intervals* of a SVR, contrasting dual and triple therapy, and CC and non-CC genotypes. When using conventional dual therapy, the *odds* for SVR are around 0.34 for non-CC genotype (corresponding to $P$ 25% probability), which increases to 1.98 ($P = 66\%$) for CC genotype. For triple therapy the *odds* are more favorable, 1.88 ($P=65\%$) for non-CC and 5.34 ($P = 84\%$) for CC genotype. The interaction effect between genotype and type of therapy is significant ($P = 0.00126$), *i.e.*, the *odds ratio* between genotypes differs between therapy types (and vice versa). According to the BIC, this model, including a treatment indicator (double *vs* triple), a genotype effect and their interaction fits the data best models that we investigated. In addition including subsets of the above variables or use a treatment indicator are also differentiating between different types of DAA.

***Comparison of individual DAA types***

In addition to the results that came out as best-fitting according to the BIC, we also analyzed the analogous results where protease inhibitor-based triple-therapy is broken down into individual subtypes (DAAs). Comparing this model and the previous one (including interactions in both cases) in an *ANOVA*, the difference between DAAs actually is significant ($P = 0.0013$). The resulting estimates are illustrated in Figure 3. Among the different DAA types, the estimated odds for SVR tend to be larger than for double therapy and greater for CC than for non-CC genotype. The only two exceptions were VNP, where SVRs for both genotypes appeared to be of the same order of magnitude and FLP, where the CC-*odds* could not be quantified. For FLP, our data originate from a single study with a 100% SVR rate (22 out of 22 patients) for CC genotype; so all we can say is that the evidence is supports effectiveness of FLP in CC genotype patients. Otherwise, for the CC genotype, the greatest odds for SVR are estimated for SMP (OR 14.71, corresponding to $P = 94\%$), whilst for non-CC genotypes, the greatest *odds* are estimated for VNP (*OR* 3.28, $P=77\%$). As in the previous model, the interaction effect between treatment type and genotype was significant ($P < 0.001$).

***Effect of patient type (IFN-$\alpha$-treatment-naïve vs IFN-$\alpha$-experienced)***

Consideration of the patient type (IFN-$\alpha$-treatment-naïve patients *vs* patients having previously experienced IFN-$\alpha$ treatment) in the regression model did not improve the model fit. Even in the best-fitting model among the ones including a patient-type effect, the patient type regarding previously IFN-$\alpha$ therapy was not significant ($P = 0.1$).

**DISCUSSION**

The main results of this meta-analysis are: I) IL28B-CC-genotype patients receiving protease inhibitor-based triple-therapy have a higher SVR rate than the IL-28B-non-CC-genotype patients with the same treatment type,

II) considering sub-types of DAAs, the effect appears to be present for BOC, FLP, SMP and TLP, but possibly not for VNP; III) IL-28B-CC-genotype patients have higher SVR rates both, in IFN-naïve and IFN experienced.

Genome-wide association studies in 2009 showed that different polymorphisms in the region of IL-28B are associated with SVR in patients chronically infected with HCV genotype-1, treated with Peg-INF-$\alpha$ and RBV[17-19]. The IL-28B gene is located on the 19 chromosome. The molecular and immunological mechanism of the IL-28B influence on SVR remains unclear[17-19]. Lately a dinucleotide polymorphism ss469415590 (TT/$\Delta$G) was described to be a better genetic predictor, as IL-28B (INF-$\lambda$3) for HCV clearance in chronically HCV genotype-1 infected patients treated with SOC[31-33]. Moreover, only the $\Delta$G of this dinucleotide polymorphism creates a novel type III interferon protein, IFN-$\lambda$4. Absence of IFN-$\lambda$4 protein is thus supposed to favor resolution of HCV infection[31,33].

The determination of IL-28B rs12979860 genotype can help to shorten the therapy duration. Genotyping of IL-28B polymorphisms can further be used to improve patient compliance, to remain on treatment in spite of side effects and to defer treatment in patients with low likelihood of response[34]. The American Association for the Study of Liver Diseases suggests IL-28B polymorphism as a robust predictive marker for treatment decision with Peg-INF-$\alpha$/RBV or in combination with DAA. Testing is useful if it impacts the treatment decision of either patient or physician. Also in studies with interferon-free therapy regimens IL-28B-CC-polymorphism was associated with better early viral kinetics and higher reduction of viral RNA[35]. Other interferon-free treatment regimens replicated these findings for IL-28B genotypes[36].

Recently, a pangenotypic polymerase inhibitor named sofosbuvir was approved in the United States of America and Europe for the treatment of chronically HCV-infected patients. In selected patients sofosbuvir achieves an SVR rate of approximately 90%. However, a 24-wk therapy with sofosbuvir and ribavirin costs about US\$ 169000[37]. Cost-effectiveness analysis show that there is no need to treat patients with IL-28B-CC allelic variation with sofosbuvir urgently because they do not necessarily benefit from such a therapy referring to the SVR rate[38]. Nevertheless, regarding the economic aspects the second-generation protease inhibitors will not be cheaper. For this reason, we need more information about predictive factors in order to detect the individuals who benefit most from an antiviral treatment with polymerase inhibitors. Through the use of predictive factors it will be possible to achieve the highest rate for SVR and the least side effects as well as reducing the cost significantly. Certainly, the IL-28B polymorphisms will play a major role in the future.

Our analyzes showed that IL-28B-CC patients could be treated with a protease inhibitors, either with FLP or SMP. Patients with IL-28B-CC who were treated with either FLP or SMP showed a SVR rate of 100% or 94%, respectively. Therefore, patients with IL-28B-CC genotype could be treated preferably with either FLP or SMP and the IL-28B-non-CC genotypes could be treated preferably with either the polymerase inhibitor sofosbuvir or with a combination of polymerase and protease inhibitors in case of an interferon-intolerance[37].

The difference between the IL-28B SNP predictive effect in triple and dual therapy is significant, suggesting that the effect of IL-28B on the *odds* of a SVR is smaller for triple-therapy than for dual-therapy.

Regarding BOC, the individual studies initially had contradictory results. The study conducted by Flamm *et al*[26] showed that for the IL-28B-TT rs12979860 genotype BOC had a favorable prognosis. However this study had a smaller number of participants than the SPRINT2 and RESPOND2 trials. Poordad, *et al*[21] analyzed the data from these studies and showed that IL-28B-CC rs12979860 genotype patients were more likely to achieve a SVR. Our analysis showed that in the case of the patients treated with BOC the CC-genotype has a favorable prognosis.

For FLP and SMP, we could only include one study each, with a relatively small number of participants. For FLP, the *odds* for the CC genotype could not be quantified, due to the fact that our data originate from a single study with 100% SVR rate, indicating a strong beneficial effect.

In the case of SMP, the IL-28B-CC-genotype has the second best *odds* among all DAAs, but with a large *CI* because of the limited number of patients that were included in the study. Therefore, future studies with these DAAs are needed to confirm these results. Our meta-analysis showed that SMP based triple therapy is more likely to produce SVR in CC-genotype patients; therefore we recommend IL-28B genotyping before initiation of this treatment.

The studies by Pol *et al*[28] and Jacobson *et al*[27] showed that IL-28B-genotype has a limited and non-significant predictive value for a SVR regarding the triple-therapy with TVR. Both of them are analyses of the data from larger trials (Pol *et al*[28] from REALIZE and Jacobson *et al*[27] from ADVANCE US) stipulating that TVR based triple-therapy increase the SVR rate through all IL-28B genotypes, especially for the IL-28B-non-CC genotype patients. In our analysis TVR based regimes included a larger number of studies ($n = 4$). The results were significantly favorable for IL-28B-CC-genotype patients. This result can be explained by the fact that Akuta *et al*[25] studied the predictive value of IL-28B-genotype only in Asian patients infected with genotype 1B, with higher SVR rates while the other studies included wider ranges of ethnicities.

IL-28B SNP has a predictive role for both, IFN-naïve and IFN-previously treated patients. For the SOC-double therapy this meta-analysis did not show evidence for a difference in treatment effect between patient types.

The strong points of our meta-analysis is the large number of patients ($n = 2707$), the included studies were randomized, controlled studies and the inclusion of various number of DAA types ($n = 5$). The limitations of our meta-analysis are the relatively small number of studies for some DAAs types (SMP, FLP), even though both, full text and meeting abstracts were included into the search. Another limitation to our study could be the absence of information on the influence of baseline viral loads on SVR and race in correlation with the IL-28B SNP. No long-term data are available yet. Furthermore, this meta-analysis reflects the methodological problems of the included studies.

In conclusion, the IL-28B allelic variation has a predictive value in the protease inhibitor-based triple-therapy of chronically HCV genotype-1 infected patients and it differs among DAA types. However, the effect on the *odds* of a SVR is smaller than the one regarding IL-28B and SOC. We recommend IL-28B genotyping also in the case of SMP-based triple therapy. VNP based regime was the only triple therapy which was not associated with higher SVR rates for IL-28B-CC-genotype patients. Furthermore, prospective studies need to be conducted for the understanding of IL-28B-genotype predictive role in HCV triple-therapy.

## ACKNOWLEDGEMENTS

We would like to thank Dr. Imke Hoell for the revision of the manuscript.

## COMMENTS

### Background

Hepatitis C is a world health problem and represents a dynamic field of research for new therapeutic options. The interleukin-28B (IL-28B) single nucleotide polymorphism (SNP) is a predictor of sustained virological response for hepatitis C genotype-1 patients treated with pegylated-Interferon-$\alpha$ and ribavirin as the standard of care. Recently, direct antiviral agents have been developed which, in addition to the standard of care, obtain higher sustained virological response rates, but with higher costs and side effects.

### Research frontiers

IL-28B is a solid genetic predictor in the therapy of hepatitis C patients treated with interferon and ribavirin. In the era of new therapeutic options for hepatitis C, the current research hotspot is to evaluate the predictive value of IL-28B in different protease inhibitor-based triple-therapies.

### Innovations and breakthroughs

This meta-analysis demonstrates that IL-28B has a predictive value on protease inhibitor-based triple-therapy. This predictability differs among protease inhibitors.

### Applications

This study suggests that IL-28B could be used as a genetic predictive factor for antiviral response in hepatitis C genotype 1 patients treated with protease inhibitor-based triple-therapy.

*Terminology*

Direct antiviral agents such as protease inhibitors are newly developed drugs against hepatitis C. In combination with Interferon and Ribavirin they constitute the triple therapy for hepatitis C. SNP within the interleukin 28B gene as a genetic marker is associated with sustained virological response in the treatment of hepatitis C.

C. *Hepatology* 2014; **59**: 1692-1705 [PMID: 24691835 DOI: 10.1002/hep.27010]

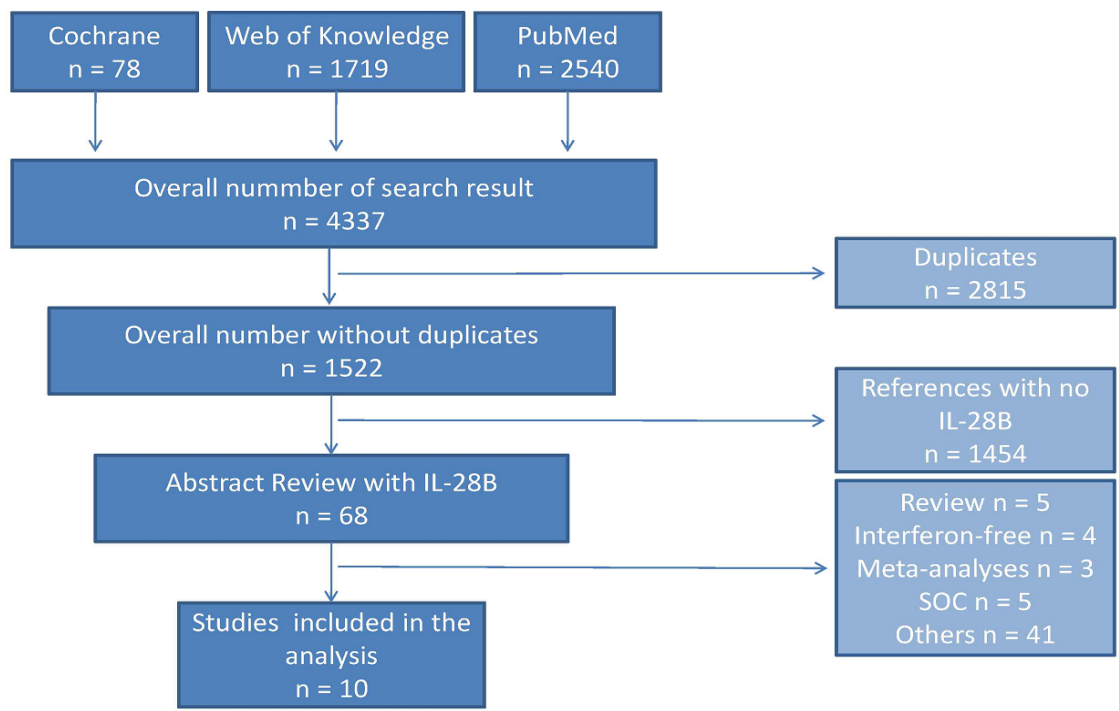

**Figure 1:** Flow chart of systematic review of protease inhibitor based triple therapy.

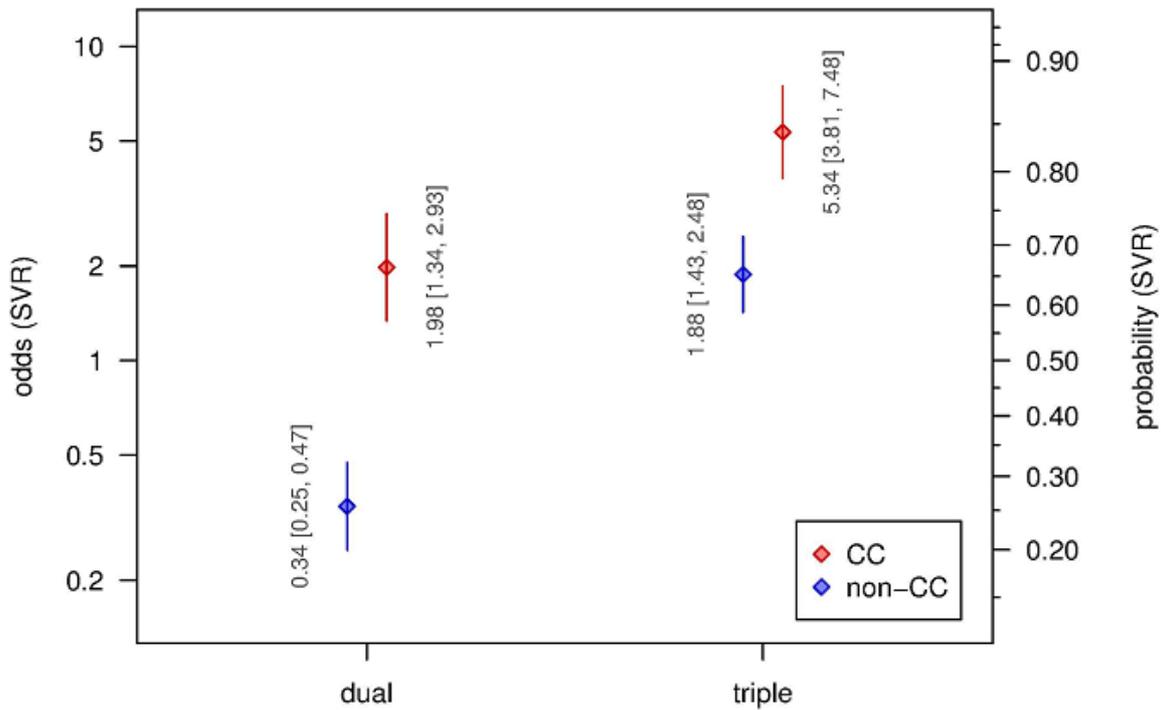

**Figure 2:** *Odds/probabilities* of obtaining a sustained virological response with regard to interleukin 28B-genotype and different therapy regimen. The differences between the shown estimates correspond to the odds ratios. A greater difference of odds between the both IL-28B-genotype corresponds to a more beneficial effect. SVR: Sustained virological response; IL-28B: Interleukin 28B.

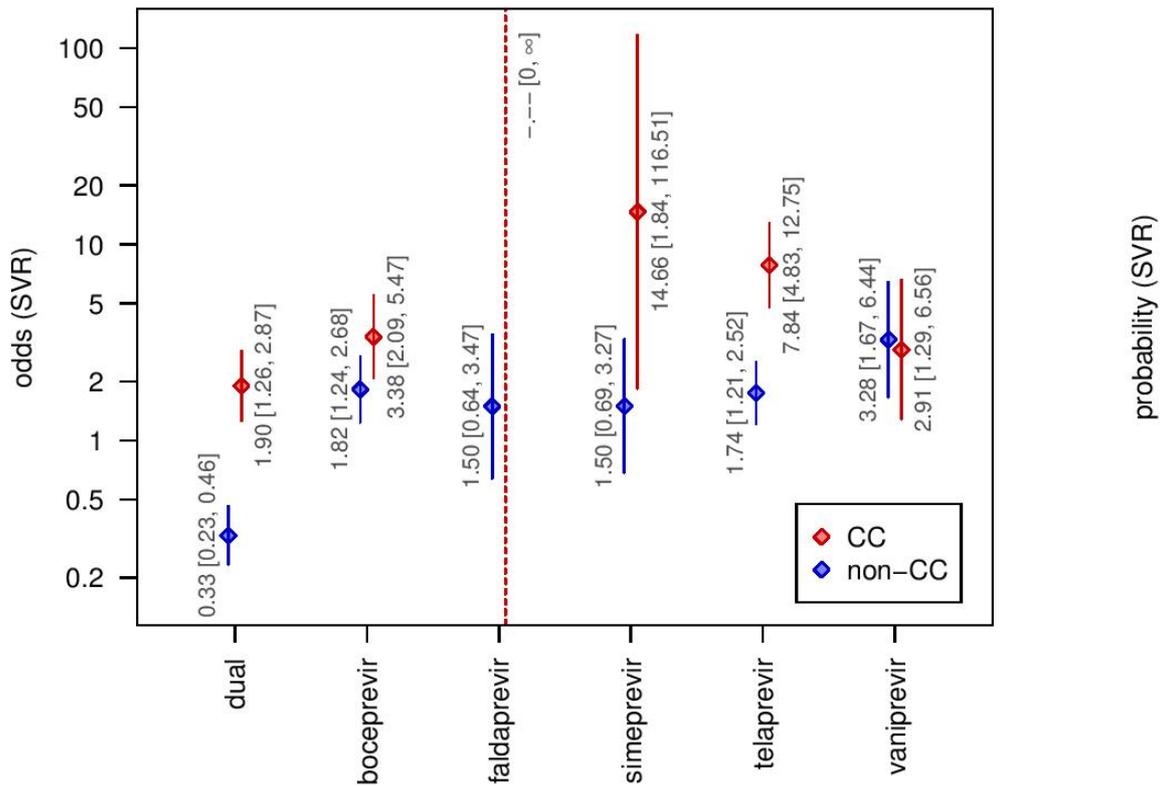

**Figure 3:** Odds/probabilities of a sustained virological response with regard to interleukin 28B-genotype in different protease inhibitor based triple therapy. The differences between the shown estimates correspond to the odds ratios. A greater difference of odds between the both IL-28B-genotype corresponds to a more beneficial effect. SVR: Sustained virological response; IL-28B: Interleukin 28B.

## Table 1 Characteristics of included trials

| Ref. | DAA type | Patient type | IL-28B SNP (n) | DAA SVR (n) CC | DAA SVR (n) Non CC | DAA Non SVR (n) CC | DAA Non SVR (n) non CC | SOC SVR (n) CC | SOC SVR (n) non CC | SOC Non SVR (n) CC | SOC Non SVR (n) non CC |
|---|---|---|---|---|---|---|---|---|---|---|---|
| Akuta et al[25] | TVR | Mixed | 68 | 31 | 10 | 6 | 21 | | | | |
| Bronowicki et al[23] | TVR | Naïve | 141 | 30 | 30 | 2 | 48 | 7 | 6 | 4 | 14 |
| Flamm et al[26] | BOC | Experienced | 146 | 12 | 52 | 7 | 24 | 5 | 6 | 5 | 35 |
| Fried et al[22] | SMP | Naïve | 153 | 34 | 56 | 1 | 16 | 1 | 17 | 0 | 17 |
| Jacobson et al[27] | TVR | Naïve | 454 | 84 | 127 | 11 | 71 | 23 | 25 | 2 | 81 |
| Lawitz et al[20] | VNP | Experienced | 131 | 14 | 67 | 9 | 16 | 51 | 3 | 02 | 19 |
| Manns et al[13] | VNP | Naïve | 65 | 22 | 14 | 3 | 10 | 4 | 6 | 1 | 5 |
| Pol et al[28] | TVR | Experienced | 527 | 60 | 209 | 16 | 137 | 5 | 13 | 1 | 75 |
| Sulkowski et al[24] | FLP | Naïve | 110 | 22 | 34 | 0 | 14 | 9 | 12 | 22 | 17 |
| Poordad et al[21] | BOC | Naïve | 653 | 107 | 198 | 25 | 106 | 50 | 43 | 14 | 110 |
| | | Experienced | 259 | 39 | 105 | 11 | 52 | 6 | 10 | 7 | 29 |

DAA: Direct acting agents; TVR: Telaprevir; BOC: Boceprevir; SMP: Simeprevir; VNP: Vaniprevir; FLP: Faldaprevir;

CC or non CC: Genotype of IL-28B.